\documentclass[prE,twocolumn,superscriptaddress,showpacs,floatfix]{revtex4-1}
\usepackage{epsfig}
\usepackage{color}
\usepackage{graphics}
\usepackage{amsmath}
\usepackage{mathrsfs}

\begin{document}

\title{Stochastic epigenetic dynamics of gene switching}

\author{Bhaswati Bhattacharyya}\email{b.bhaswati@gmail.com}
\affiliation{Department of Applied Physics, Nagoya University,
Nagoya 464-8603, Japan}

\author{Jin Wang}\email{jin.wang.1@stonybrook.edu}
\affiliation{Department of Chemistry, Physics and Applied Mathematics, State University of New
York at Stony Brook, Stony Brook, New York, 11794-3400, USA}

\author{Masaki Sasai}\email{masakisasai@nagoya-u.jp}
\affiliation{Department of  Applied Physics, Nagoya University,
Nagoya 464-8603, Japan}

\date{\today}

\begin{abstract}
Epigenetic modifications of histones crucially affect the eukaryotic gene activity, while the epigenetic histone state is largely determined by the binding of  specific factors such as the  transcription factors (TFs) to DNA. Here, the way how the TFs and the histone state are dynamically correlated is not obvious when the TF synthesis is regulated by the histone state. This type of feedback regulatory relations are ubiquitous in gene networks to determine cell fate in differentiation and other cell transformations.
To gain insights into such dynamical feedback regulations, we theoretically analyze a model of epigenetic gene switching by extending the Doi-Peliti operator formalism of reaction  kinetics to the problem of coupled molecular processes. The spin-1 and spin-1/2 coherent state representations are introduced to describe stochastic reactions of histones and binding/unbinding of TF in a unified way, which provides a concise view of the effects of timescale difference among these molecular processes; even in the
case that binding/unbinding of TF to/from DNA are adiabatically fast, the slow nonadiabatic histone dynamics give rise to a distinct circular flow of the probability flux around basins in the landscape of the gene state distribution, which leads to hysteresis in gene switching. In contrast to the general belief that the change in the amount of TF precedes the histone state change, the flux drives histones to be modified prior to the change in the amount of TF in the self-regulating circuits. The flux-landscape analyses shed light on the nonlinear nonadiabatic mechanism of epigenetic cell fate decision making.
\end{abstract}

\maketitle

\section{Introduction}
Epigenetic pattern formation associated with chemical modifications of histones provides long-term memory of gene regulation, which plays a critical role in cell differentiation, reprogramming, and oncogenesis \cite{Bonasio2010,Flavahan2017}. The effects of epigenetic modifications have been discussed theoretically \cite{Sasai2013,Ashwin2015,Feng2011,Feng2012,Li2013,Chen2013,Chen2016,Folguera2019,Tian2016,Yu2019}; however, their quantitative dynamics still remain elusive. Statistical mechanical analyses have suggested that histones in an array of $\sim 10^2$--$10^3$ interacting nucleosomes (i.e., particles of histone-DNA complex) show collective changes in their modification pattern \cite{Dodd2007,Zhang2014,Sood2020}, and such collective modifications have been indeed found in loops or domains of chromatin \cite{Dixon2012,Rao2014,Boettiger2016}. It has been traditionally considered that concentration of transcription factor (TF) largely determines the histone state \cite{Ptashne2013}; however, the relation between TF and the collective histone state is not obvious when both TF and the histone state are regulated with a network of feedback loops. In particular, the mechanism of how histone modifications are induced prior to the gene activation in developmental processes remains a mystery \cite{Samata2020,Sato2019,Zhang2018}. Therefore, to get physical insights into these dynamical regulatory systems, the relation between TF and the histone state needs to be tested by physical models.

In physical modelling, particularly important  is to represent the collective histone state as a dynamically fluctuating variable to examine the effects of TF binding/unbinding on the histone state dynamics. Here, we need to take account of  the effects of timescale difference among various molecular processes; when a certain molecular process is much faster than the other processes, the fast process can be regarded as being averaged as in quasi-equilibrium in the adiabatic approximation, while the slowest process could be regarded as  stationary in the nonadiabatic limit. The effects of adiabaticity/nonadiabaticity on the simple bacterial gene dynamics have been intensively investigated from the statistical physical viewpoint with theoretical \cite{Hornos2005,Walczak2005b,Schutz2007,Yoda2007,Okabe2007,Feng2011,Shi2011,Zhang2013,Chen2015} and experimental \cite{Jiang2019,Fang2018} methods, which revealed the large fluctuation of gene activity in the middle-range regime between  adiabatic and nonadiabatic limits. However, the effects on the more complex eukaryotic genes have remained challenging \cite{Sasai2013,Zhang2014b,Ashwin2015,Yu2019}. In the present paper, we investigate the problem of adiabaticity/nonadiabaticity in eukaryotic genes by explicitly taking account of the degree of collective  histone state transitions, which are the mechanism absent from bacterial genes but play a central role in eukaryotic gene switching.

A straightforward way to analyze the physical models of gene regulation  is to  simulate their kinetics with a Monte-Carlo type numerical method. However, such a calculation does not necessarily provide a global understanding and physical picture directly. A clearer picture would be obtainable when the stochastic kinetics is described with  the chemical Langevin equation, which emerges in the high molecular number or large volume limit as a continuous description of stochastic chemical reactions. The Langevin dynamical method leads to the combined description of probability flux and landscape, where the landscape represents the distribution of states generated via nonlinear interactions and the flux reflects the nonequilibrium driving force of transitions among states. 
 The combined flux-landscape analyses have been  useful to obtain a global and physical perspective of various complex systems \cite{Fang2019,Wang2008,Wang2011}; however, in the present problem of gene switching, the continuous  changes of molecular concentrations  are coupled with the discrete processes of TF binding/unbinding  and transitions of histone states. Therefore, to describe such coupled discrete and continuous dynamics, 
we need to consider the coupled multiple landscapes; the Langevin dynamics are the motions on individual landscapes and the discrete molecular state changes are transitions among  landscapes \cite{Zhang2013,Chen2015}. In this problem, even utilizing the flux-landscape method on individual landscapes, it is still difficult to obtain a global picture when the  local stochastic transitions among multiple landscapes are frequent.  To overcome this difficulty, 
we here develop a theoretical method  by extending the Doi-Peliti operator formalism \cite{Doi1976,Peliti1985,Mattis1998} of chemical reaction kinetics. By introducing the spin-1 and spin-1/2 coherent-state representations, the combined discrete and continuous description is transformed to a unified continuous description with expanded dimension. Then, the coupled molecular processes are described as continuous dynamics on a single landscape in the higher dimensional space, which leads to a global picture and quantification of gene switching dynamics. Using this extended Doi-Peliti method, we show that the nonadiabatic dynamics of histone state transitions  give rise to a nontrivial temporal correlation and hysteresis in eukaryotic gene switching.

\section{A physical model of eukaryotic gene switching}
\subsection{Self-regulating genes}
Since the histones modified with the gene-repressing marks  and the histones having the gene-activating marks  are not directly transformed to each other but are transformed through the multiple steps of the mark-deletion and the mark-addition or the replacement with the newly synthesized unmarked histones, it is natural to describe the individual histones with three states; gene-repressing, unmarked, and gene-activating states \cite{Dodd2007,Zhang2014}. Similarly, the quantitative experimental data of  transitions among   collective histone states, which emerge through the cooperative interactions among many nucleosomes, have been fitted by the three-state transition models \cite{Hathaway2012,Bintu2016}. Therefore, in the similar way to those previous models, we classify the collective histone states around a promoter site of a modeled gene into three states: the gene-activating state with histones marked as H3K9ac or others ($s=1$), the gene-repressing state marked as H3K9me3 or others ($s=-1$), and the neutral state with mixed or an absence of activating and repressing marks ($s=0$). We write the transition rate from  state $s'$ to $s$ as $r_{s s'}$. The chromatin chain with $s=1$ takes an open structure, which facilitates access of the large-sized transcribing molecules to DNA to enhance the gene activity, whereas the $s=-1$ chromatin is condensed, which prevents the access of necessary molecules to suppress the gene activity (Fig.~1).

We assume that the gene activity is regulated by both the histone state and binding of a TF; TF binds to DNA near the promoter site ($\sigma =1$) or unbinds from DNA ($\sigma =-1$) with binding rate $h$ and unbinding rate $f$.  
These rates are insensitive to the chromatin packing density when the size of TF is as small as the pioneer TFs which can diffuse into the compact chromatin space \cite{Soufi2012,Chen2014}. Here, for simplicity, we consider  that the TF is a pioneer factor as Sox2 or Oct4 in mammalian cells \cite{Chen2014} and its binding/unbinding rates are not affected significantly by the chromatin compactness or the histone state. However, the bound TF should recruit histone modifier enzymes, so that binding of a single TF  nucleates the histone state change, which is expanded and propagates along the DNA sequence to induce the collective histone state change  as observed in engineered  \cite{Hathaway2012,Bintu2016,Park2019} and  model cells \cite{Hall2002,Rusche2002}.  Therefore, binding of the TF modifies $r_{ss'}$ as $r_{s s'}=r_{s s'}^0+\delta_{\sigma 1}\Delta r_{s s'}$, where $\delta_{\sigma 1}$ is a Kronecker delta. The rate of collective change in many histones, $r_{s s'}$, should be smaller in general than the rate of single-molecule binding/unbinding, $h$ or $f$, which enables histones to retain the effect of TF-binding as memory; $\Delta r_{s s'}> 0$ for $s>s'$ when the TF is an activator while $\Delta r_{s s'}> 0$ for $s<s'$ when the TF is a repressor.

%%%%%%%%%%%%%%%%%%%%%%%%%%%%%%%%%%%%%%%%%%%%%
\begin{figure}[t]
\begin{center}
\includegraphics[width=8.5cm]{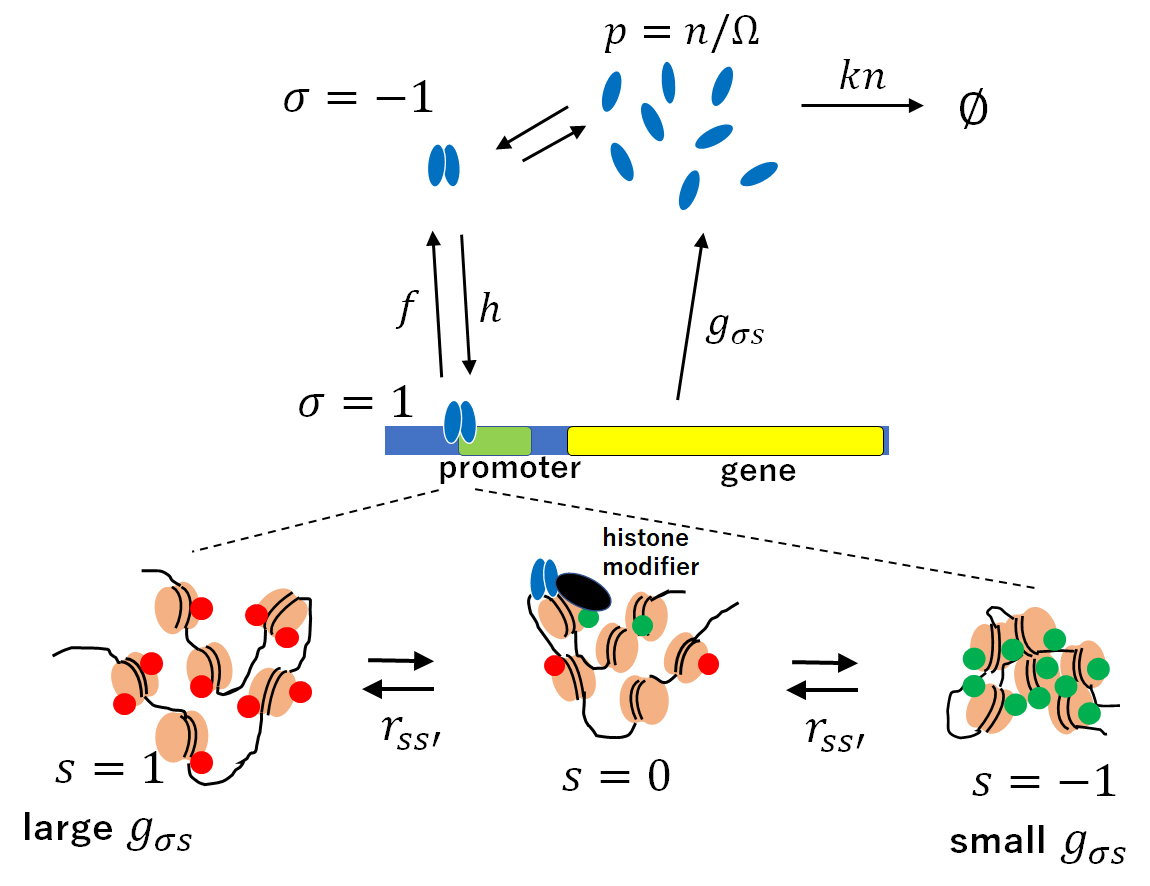}
\caption{
Schematic of a self-regulating gene. Close-up views of chromatin at around the promoter region are illustrated at the bottom, where histones (brown) are actively marked (red) or repressively marked (green). Histones in the chromatin are collectively marked through their mutual interactions. When the activating mark is dominant, chromatin has an open structure, which enhances the gene transcription activity ($s=1$), and when the repressive mark is dominant, condensed chromatin suppresses the gene activity ($s=-1$). When neither mark is dominant, the histone state is neutral ($s=0$). The transcription factor (TF) is a dimer of the product protein (blue oval). The bound TF recruits histone modifiers (black), which trigger the collective histone modification as was observed in engineered cells \cite{Hathaway2012,Bintu2016,Park2019} by perturbing the transition rates $r_{s s'}$ among the histone states. The protein production rate $g_{\sigma s}$ depends both on the TF binding status, $\sigma$, and on the histone state, $s$. See the text for the definition of rates denoted on the arrows. }
\label{fig_model}
\end{center}
\end{figure}
%%%%%%%%%%%%%%%%%%%%%%%%%%%%%%%%%%%%%%%%%%%%%

As a prototypical motif of gene circuits, we consider a self-regulating gene as in Fig.~1; a dimer of the product protein is the TF acting on the gene itself. Here, by assuming that dimerization is much faster than the other reactions, the TF-binding rate should be $h=h_0p^2$, where $p=n/\Omega$ is the protein concentration, $n$ is the protein copy number in the nucleus, 
$\Omega$ is a typical copy number of the protein in the nucleus when the transcription from the gene is active,
 and $h_0$ is a constant. The protein production rate depends on both $\sigma$ and $s$ as $g_{\sigma s}$. Here, $g_{11}$ is the largest and $g_{-1-1}$ is the smallest of $g_{\sigma s}\ge 0$ for an activator TF, whereas $g_{-11}$ is largest and $g_{1-1}$ is smallest for a repressor TF. With the approximation that the protein degradation depends on the total copy number $n$, the degradation rate is $kn$ with a constant $k$. In this way, the rates of the histone state transitions are affected by the binding status of the TF, while the binding rate of the TF is regulated by the amount of the TF, and the synthesis rate of the TF is affected by the histone state. Therefore, the histone state is a part of the feedback loop of the regulation. Because the rates of the histone state transitions are smaller than the TF binding/unbinding rates in general, the histone state constitutes a relatively  slowly varying part in the feedback loop.

Self-activating motifs discussed in the present paper are ubiquitous in cells. In mouse embryonic stem cells (mESCs), for example,  the genes necessary for sustaining pluripotency are activating each other. A Sox2-Oct4 heterodimer binds on the gene loci of {\it Sox2} and {\it Oct4} and activates themselves \cite{Okumura2005,Masui2007}. The self-activator gene in the present paper can be regarded as  a simplified model of this {\it Sox2}-{\it Oct4} system when these two genes are described  as the strongly correlated loci.

\subsection{Operator formalism of reaction kinetics}

To describe the reactions in Fig.~1, we use the operator formalism of Doi and Peliti \cite{Doi1976,Peliti1985,Mattis1998}, which has been applied to the problems of gene regulation without explicitly considering the histone state $s$ \cite{Sasai2003,Walczak2005,Ohkubo2008,Zhang2013,Zhang2014b} and to the problem of histone state without considering the TF-binding status $\sigma$ \cite{Sood2020}. Now, taking into account those processes having different timescales in a unified way, we consider the probability that the protein copy number is $n$ at time $t$, $P_{\sigma s}(n,t)$. We define a six dimensional vector ${\boldsymbol \psi} (t)$, whose component is ${\boldsymbol \psi} (t)_{\sigma s}=\sum_n P_{\sigma s}(n,t)|\left. n\right>$. The operators $a$ and $a^{\dag}$ are introduced as $a^{\dag}|\left. n\right>=|\left. n+1\right>$, $a|\left. n\right>=n|\left. n-1\right>$, and $[a,a^{\dag}]=1$. Then, by assuming that all the reactions are Markovian, the master equation of the reactions is 
$\frac{\partial }{\partial t}{\boldsymbol \psi} =-\mathscr{H}{\boldsymbol \psi}$,
with $\mathscr{H}$ being a six dimensional operator, 
\begin{equation}
\mathscr{H}= k(a^{\dag}a-a){\bf 1}+G(1-a^{\dag})+J\otimes K, 
\end{equation}
where {\bf 1} is a unit matrix, $G$ is a diagonal matrix, $G_{\sigma s, \sigma' s'}=\delta_{\sigma s, \sigma' s'}\Omega g_{\sigma s}$, and $J$ and $K$ are transition matrices for $\sigma$ and $s$, respectively;
\begin{eqnarray}
J &=&
\left( \begin{array}{cc}
-f & h \\
f & -h \\
\end{array} \right), \nonumber \\
K &=&
\left( \begin{array}{ccc}
-r_{01} & r_{10} & 0\\
r_{01} & -r_{10}-r_{-10} & r_{0-1}\\
0 & r_{-10} & -r_{0-1}\\ 
\end{array} \right),
\end{eqnarray} 
with $h=h_0(a^{\dag}a)^2$.

We should note that $\mathscr{H}$ is non-Hermitian reflecting the nonequilibrium features of the processes in gene expression.
From Eq.~1, we can formally write the transition probability matrix ${\bf P}(n_f,\tau | n_i,0)$ between  the state $n=n_i$ at $t=0$ and the state $n=n_f$ at $t=\tau$  as
\begin{equation}
{\bf P}(n_f,\tau | n_i,0)=\frac{1}{n_f!}\left< n_f| \right. \exp \left( -\int_0^{\tau}dt \mathscr{H} \right) \left. |n_i\right>.
\end{equation}

\subsection{Continuous dynamics in the higher dimensions}
The temporal development of ${\boldsymbol \psi}(t)$ can be represented in a path-integral form by using the coherent-state representation, $|\left.z\right>=e^{a^{\dag}z}|\left. 0\right>$, with a complex variable $z(t)$. 
The transition paths in the $\sigma$ and $s$ space can be represented by using the spin-1/2 and spin-1 coherent states with spin angles $\theta(t)$ and $\alpha(t)$ and their conjugate variables $\phi (t)$ and $\beta (t)$;
\begin{eqnarray}
{\boldsymbol \sigma}({\theta, \phi})&=&e^{i\phi/2}(\cos^2\frac{\theta}{2}){\boldsymbol \sigma}_1+e^{-i\phi/2}(\sin^2\frac{\theta}{2}){\boldsymbol \sigma}_{-1} \nonumber \\
{\bf s}({\alpha, \beta})&=&e^{i\beta}(\cos^4\frac{\alpha}{2}){\bf s}_1+2(\cos^2\frac{\alpha}{2}\sin^2\frac{\alpha}{2}){\bf s}_0  
\nonumber \\
&+&e^{-i\beta}(\sin^4\frac{\alpha}{2}){\bf s}_{-1},
\end{eqnarray}
with ${\boldsymbol \sigma}_i=(\delta_{i1},\delta_{i-1})^T$ for TF binding/unbinding
and  ${\bf s}_i=(\delta_{i1},\delta_{i0},\delta_{i-1})^T$ for the histone degree of freedom.
Considering the non-Hermiticity of $\mathscr{H}$, we introduce the conjugate vectors, 
\begin{eqnarray}
\left<z\right. | &=& \left<0 \right. |e^{az^*}, \nonumber \\
\tilde{\boldsymbol \sigma} &=& e^{-i\phi/2}{\boldsymbol \sigma}_1^T+e^{i\phi/2}{\boldsymbol \sigma}_{-1}^T, \nonumber \\
\tilde{\bf s} &=& e^{-i\beta}{\bf s}_1^T+{\bf s}_0^T+e^{i\beta}{\bf s}_{-1}^T,
\end{eqnarray}
where $z=\psi e^{-i\chi}$, $\psi=|z|$, and $z^*=e^{i\chi}$.  With pairs of vectors defined in Eqs.~4 and 5, we have the identity matrix ${\bf 1}=I_z\otimes I_{\sigma}\otimes I_{s}$ as
\begin{eqnarray}
I_z &=& \frac{1}{2\pi}\int_0^{\pi}d\psi\int_{-\pi}^{\pi}d\chi | z\left>\right< z|e^{-\psi}, \nonumber \\
I_{\sigma} &=& \frac{1}{2\pi}\int_0^{\pi}\sin\theta d\theta \int_0^{2\pi}d\phi
\left( \tilde{\boldsymbol \sigma}{\boldsymbol \sigma} \right), \nonumber \\
I_{s} &=& \frac{3}{4\pi}\int_0^{\pi}\sin\alpha d\alpha\int_0^{2\pi}d\beta \left(\tilde{\bf s}{\bf s}\right).
\end{eqnarray}
Using Eq.~6, Eq.~3 is represented in a path-integral form as
\begin{eqnarray}
&&{\bf P}(n_f,\tau | n_i,0) \nonumber \\
&&=const.\int D\alpha D\beta D\theta D\phi D\psi D\chi \exp\left( -\int \mathscr{L} dt \right),
\end{eqnarray} 
where $\mathscr{L}$ is an effective ``Lagrangian'', 
$\mathscr{L} = \mathscr{L}_{\chi\psi}+\mathscr{L}_{\phi\theta}+\mathscr{L}_{\beta\alpha}$, and
\begin{eqnarray}
\mathscr{L}_{\chi\psi} =&& i\chi\frac{d\psi}{dt} 
+\Omega(1-e^{i\chi})\left[ g_{-1}(\theta)\sin^4\frac{\alpha}{2} \right. \nonumber \\
&& \left. 
+2g_0(\theta)\sin^2\frac{\alpha}{2}\cos^2\frac{\alpha}{2}  
 +g_1(\theta)\cos^4\frac{\alpha}{2}\right]  \nonumber \\
&& +(1-e^{-i\chi})k\psi , 
\nonumber \\
\mathscr{L}_{\phi\theta} =&& i\frac{\phi}{2}\frac{d}{dt}\cos\theta 
\nonumber \\
&&+h_0\Omega^{-2}\psi^2(1-e^{i\phi})\sin^2\frac{\theta}{2}+f(1-e^{-i\phi})\cos^2\frac{\theta}{2} , 
\nonumber \\
\mathscr{L}_{\beta\alpha} =&& i\beta\frac{d}{dt}\cos\alpha 
\nonumber \\
&&+(1-e^{i\beta})\left[r_{0-1}(\theta)\sin^4\frac{\alpha}{2}+2r_{10}(\theta)\sin^2\frac{\alpha}{2}\cos^2\frac{\alpha}{2}\right]
\nonumber \\
&&+(1-e^{-i\beta})\left[2r_{-10}(\theta)\sin^2\frac{\alpha}{2}\cos^2\frac{\alpha}{2} \right. \nonumber \\
&& \left. +r_{01}(\theta)\cos^4\frac{\alpha}{2}\right] ,
\end{eqnarray}
with $g_s(\theta)=g_{1s}\cos^2\frac{\theta}{2}+g_{-1s}\sin^2\frac{\theta}{2}$  and
$r_{ss'}(\theta)=r_{ss'}^0+\Delta r_{ss'}\cos^2\frac{\theta}{2}$.

Then, by retaining up to the 2nd order terms of $\chi$, $\phi$, and $\beta$ in Eq.~7 (i.e., using the saddle-point approximation) in a similar way to the method in \cite{Zhang2013}, we obtain the Langevin equations describing fluctuations in the protein concentration $p=\psi/\Omega$, the TF binding status $\xi=\cos\theta$, and the histone state $\zeta=\cos\alpha$; 
\begin{eqnarray}
\frac{dp}{dt}&=&F_p^{+}(\theta, \alpha)-kp +\eta_p, \nonumber \\
\frac{d\xi}{dt}&=&2h_0p^2\xi_{\rm s}-2f\xi_{\rm c} +\eta_{\xi}, \nonumber \\
\frac{d\zeta}{dt}&=&F_{\zeta}^{+}(\theta, \alpha)-F_{\zeta}^{-}(\theta, \alpha) +\eta_{\zeta},
\end{eqnarray}
where
\begin{eqnarray}
F_p^{+}&=&\xi_{\rm c}\left[ g_{11}\zeta_{\rm c}^2+2g_{10}\zeta_{\rm c}\zeta_{\rm s} +g_{1-1}\zeta_{\rm s}^2 \right] \nonumber \\
&+&
\xi_{\rm s}\left[ g_{-11}\zeta_{\rm c}^2+2g_{-10}\zeta_{\rm c}\zeta_{\rm s} +g_{-1-1}\zeta_{\rm s}^2 \right], \nonumber \\
F_{\zeta}^{+}&=&2\left( r_{10} +\xi_{\rm c}\Delta r_{10} \right)\zeta_{\rm c}\zeta_{\rm s}
+\left(r_{0-1}+\xi_{\rm c}\Delta r_{0-1}\right)\zeta_{\rm s}^2, \nonumber \\
F_{\zeta}^{-}&=&\left(r_{01}+\xi_{\rm c}\Delta r_{01}\right)\zeta_{\rm c}^2 
+2\left( r_{-10} +\xi_{\rm c}\Delta r_{-10} \right)\zeta_{\rm c}\zeta_{\rm s}, 
\end{eqnarray}
with $\xi_{\rm c}=\cos^2(\theta/2)$, $\xi_{\rm s}=\sin^2(\theta/2)$, $\zeta_{\rm c}=\cos^2(\alpha/2)$, and $\zeta_{\rm s}=\sin^2(\alpha/2)$. In Eq.~9, $\eta_p$, $\eta_{\xi}$, and $\eta_{\zeta}$ are mutually independent Gaussian noises with $\left<\eta_x\right>=0$, $\left<\eta_x(t)\eta_x(t')\right>=B_x\delta(t-t')$ for $x=p$, $\xi$, or $\zeta$ as
\begin{eqnarray}
B_p&=&\left( F_p^{+}(\theta, \alpha)+kp\right)/\Omega , \nonumber \\
B_{\xi}&=&4h_0p^2\xi_{\rm s}+4f\xi_{\rm c}, \nonumber \\
B_{\zeta}&=&F_{\zeta}^{+}(\theta, \alpha)+F_{\zeta}^{-}(\theta, \alpha).
\end{eqnarray}
We should note that $p=n/\Omega$ was an almost continuous variable for a large $\Omega$. Hence, the original coupled dynamics of a nearly continuous variable $p$ and the discrete variables, $\sigma$ and $s$, were transformed into the Brownian dynamics in the 3-dimensional (3D) space of the continuous variables, $p$, $\xi$, and $\zeta$, with TF bound ($\xi=1$)/unbound ($\xi=-1$) and histone state activating ($\zeta=1$)/neutral ($\zeta=0$)/repressing ($\zeta=-1$).

In the numerical calculations of Eq.~9, infrequent but  large noises may push  $p$, $\xi$, and $\zeta$  outside the originally defined range of  values, $p\ge 0$, $-1\le \xi\le 1$, and $-1\le \zeta \le 1$.
In order to reduce this out-of-range fluctuations, we add soft-wall forces, $w_p$, $w_{\xi}$, and $w_{\zeta}$,  to  Eq.~9 in numerical simulations as $\frac{dp}{dt}=F_p^{+}(\theta, \alpha)-kp +w_p+\eta_p$, 
$\frac{d\xi}{dt}=2h_0p^2\xi_{\rm s}-2f\xi_{\rm c} +w_{\xi}+\eta_{\xi}$, and
$\frac{d\zeta}{dt}=F_{\zeta}^{+}(\theta, \alpha)-F_{\zeta}^{-}(\theta, \alpha) +w_{\zeta}+\eta_{\zeta}$ with
\begin{eqnarray}
w_p &=& \left\{ \begin{array}{cc}
-c(p-0.1)^3 & {\rm for}\,\,\, p\le 0.1 \\
0 & {\rm otherwise},
\end{array} \right. \nonumber \\
w_{\xi} &=& \left\{ \begin{array}{cc}
-c(\xi+0.9)^3 & {\rm for}\,\,\, \xi\le -0.9 \\
-c(\xi-0.9)^3 & {\rm for}\,\,\, \xi\ge 0.9 \\
0 & {\rm otherwise},
\end{array} \right. \nonumber \\
w_{\zeta} &=& \left\{ \begin{array}{cc}
-c(\zeta+0.9)^3 & {\rm for}\,\,\, \zeta\le -0.9 \\
-c(\zeta-0.9)^3 & {\rm for}\,\,\, \zeta\ge 0.9 \\
0 & {\rm otherwise},
\end{array} \right.
\end{eqnarray}
with a constant $c>0$.

\subsection{Adiabatic approximation of TF binding/unbinding}

In a single-molecule tracking experiment of the TF in mESCs, the observed timescale of binding/unbinding was $\sim 1/{\rm min}$ \cite{Chen2014}, whereas the observed degradation rate $k$ of TF was $\sim 0.1/{\rm hour}$ \cite{Chew2005,Thomson2011}. Though a single histone can be replaced in $\sim {\rm hour}$, many histones collectively change with the rate $\sim 1/{\rm day}$ \cite{Hathaway2012}. Therefore, the estimated ratios are $h/k\sim f/k=O(10^2)$ and $r_{ss'}/k=O(1)\sim O(10^{-1})$. With such large $h$ and $f$, TF-binding/unbinding reactions can be treated as {\it adiabatic}: TF-binding/unbinding are regarded as in quasi-equilibrium as $\left<d\xi/dt\right>=0$, leading to $\xi_{\rm s}=1/\left[(h_0/f)p^2+ 1\right]$. With this adiabatic approximation, the 3D calculation in Eq.~9 for $(p,\xi,\zeta)$ is reduced to the 2D calculation for $(p,\zeta)$. On the other hand, the rate of collective histone change is small; hence, the histone dynamics remains {\it nonadiabatic}.

Validity of the adiabatic treatment of TF binding/unbinding can be checked by comparing the simulated results of the present model with the experimental data of Hathaway et al. \cite{Hathaway2012}. Hathaway et al. developed a technique to forcibly bind a chromo-shadow domain of HP1 (csHP1) to a DNA site near the promoter of {\it Oct4} in mESCs. The bound csHP1 nucleated the repressively marked histones and the histones around the promoter region were collectively transformed to the repressive state in several days. In Fig.~2, this observation is compared with the calculated results obtained by applying the adiabatic approximation of TF binding/unbinding to Eq.~9. We assumed that csHP1 binding intensively suppresses the acetylation and other activating modifications of histones, so as to reduce the corresponding $r_{ss'}/k$ from $O(10^{-1})$ to $O(10^{-2})$. The simulated results reproduce the relaxation of the collective histone state toward the repressive state after csHP1 binding. Here, the experimental histone state was quantified as $\zeta(t) = H_{\rm active}(t)/H_{\rm active}(0)-H_{\rm repress}(t)/H_{\rm repress}({\rm day}5)$, where $H_{\rm active}(t)$ and $H_{\rm repress}(t)$ are the observed fractions of histones marked as H3K27ac  and  H3K9me3, respectively, in the $\sim 10$~kb region around the {\it Oct4} gene.  {\it Oct4} was transcriptionally active  for $t<0$, but its histone state was modified with the bound csHP1  for $t\ge 0$. Fig.~2 shows that the adiabatic approximation of TF binding/unbinding reasonably describes the relaxation of the system to the repressive state when the slow nonadiabatic histone dynamics are suitably  assumed.

%%%%%%%%%%%%%%%%%%%%%%%%%%%%%%%%%%%%%%%%%%%%%
\begin{figure}[t]
\begin{center}
\includegraphics[width=6cm]{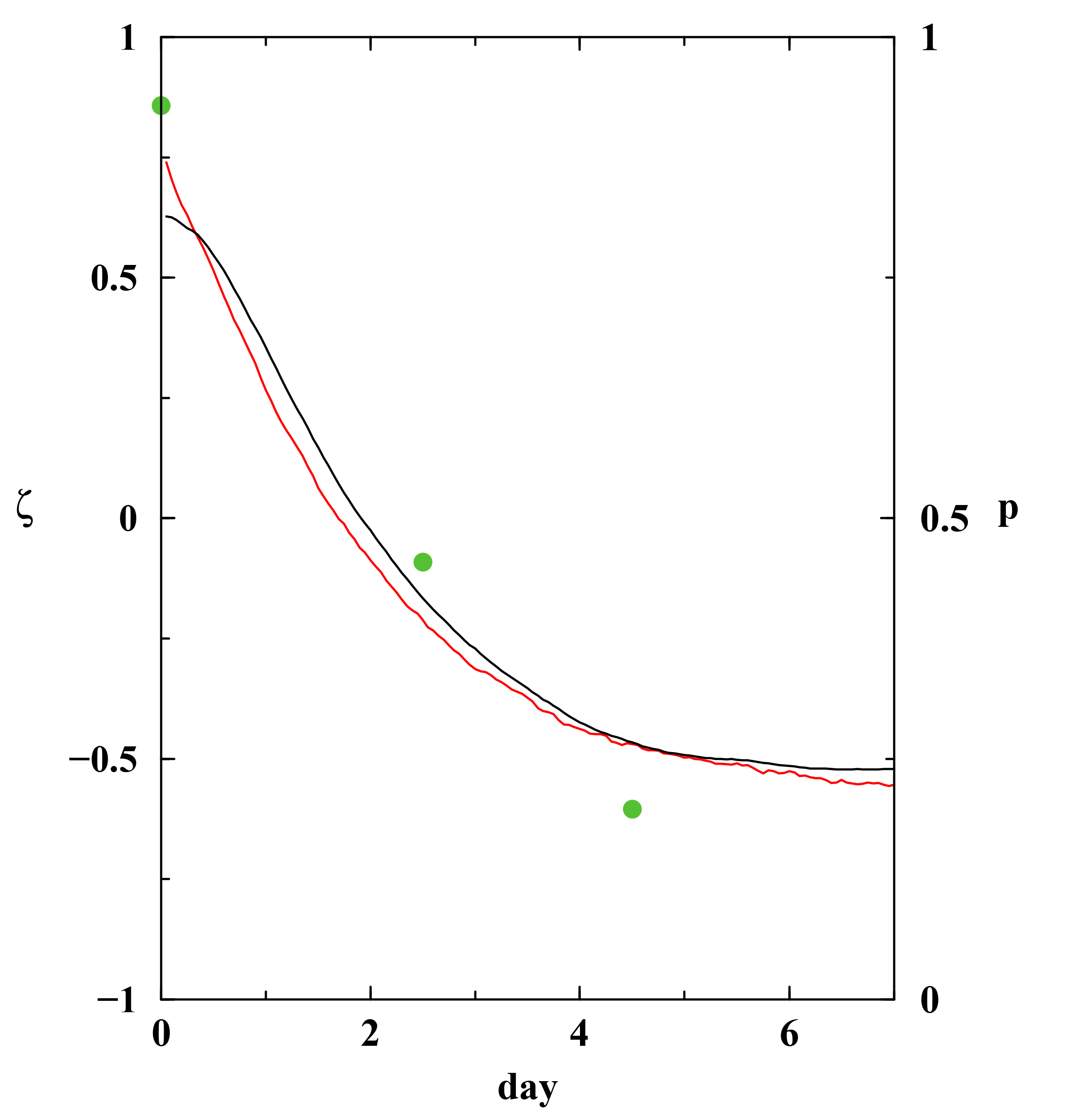}
\caption{Relaxation of a self-activating gene to the repressive state. Eq.~9 was numerically calculated by applying the adiabatic approximation of TF binding/unbinding. The protein concentration $p(t)$ (black line) and the histone state $\zeta (t)$ (red line) were derived by averaging 1000 simulated trajectories.
See text for the experimental estimation of $\zeta(t)$ (green dots) from the data of \cite{Hathaway2012}.
$k$ was set to $k=2~{\rm day}^{-1}$, and the other rates were defined in units of $k$; for $t<0$, 
 $r_{10}^0=r_{0-1}^0=0.7$, $r_{01}^0=r_{-10}^0=0.1$, and $\Delta r_{10}=\Delta r_{0-1}=1$, and
for $t\ge 0$, $r_{10}^0=r_{0-1}^0=r_{01}^0=0.01$, $r_{-10}^0=0.7$, and  $\Delta r_{10}=\Delta r_{0-1}=0.01$. For both $t<0$ and $t\ge 0$, 
$\Delta r_{01}=\Delta r_{-10}=0$, $h_0/f=200$, 
$g_{11}=1$, $g_{-11}=g_{10}=0.2$, 
$g_{-10}=g_{1-1}=g_{-1-1}=0$, 
$\Omega =100$, and $c=20$.
} 
\label{fig_landscape}
\end{center}
\end{figure}
%%%%%%%%%%%%%%%%%%%%%%%%%%%%%%%%%%%%%%%%%%%%%

\section{Landscapes, circular fluxes, and temporal correlation}

Either with adiabatic or nonadiabatic treatment of TF-binding/unbinding kinetics, the flux-landscape approach \cite{Fang2019,Wang2011,Zhang2013,Chen2015} provides a concise view of Eq.~9. With the adiabatic approximation for example, the landscape, $U(p,\zeta)$, is obtained from the calculated stationary probability density distribution as $U(p,\zeta)=-\ln P(p,\zeta)$. The probability flux ${\bf J}$ is obtained as a 2D vector field in the adiabatic TF-binding/unbinding case from the Fokker-Planck equation corresponding to Eq.~9, $\frac{\partial}{\partial t}P(p,\zeta,t)=-\nabla \cdot {\bf J}(p,\zeta,t)$, where $\nabla = (\partial_p, \partial_{\zeta})$ and ${\bf J}=(J_p, J_{\zeta})$ with $J_p=[ F_p^{+}(\theta, \alpha)-kp +w_p]P-\frac{1}{2}\frac{\partial}{\partial p}[ B_pP ]$ and $J_{\zeta}=[ F_{\zeta}^{+}(\theta, \alpha)-F_{\zeta}^{-}(\theta, \alpha) +w_{\zeta}]P-\frac{1}{2}\frac{\partial}{\partial \zeta}[ B_{\zeta}P ]$. Of note, even in the stationary state, ${\bf J}$ can be nonzero when $\nabla\cdot {\bf J}=0$. This divergence-less circulating flux is a hallmark of the breaking of the detailed balance reflecting the biased thermal/chemical energy flow such as the nucleotide consumption in protein synthesis and the heat dissipation \cite{Wang2008,Wang2011,Zhang2013}.

%%%%%%%%%%%%%%%%%%%%%%%%%%%%%%%%%%%%%%%%%%%%%
\begin{figure}[t]
\begin{center}
\includegraphics[width=8.5cm]{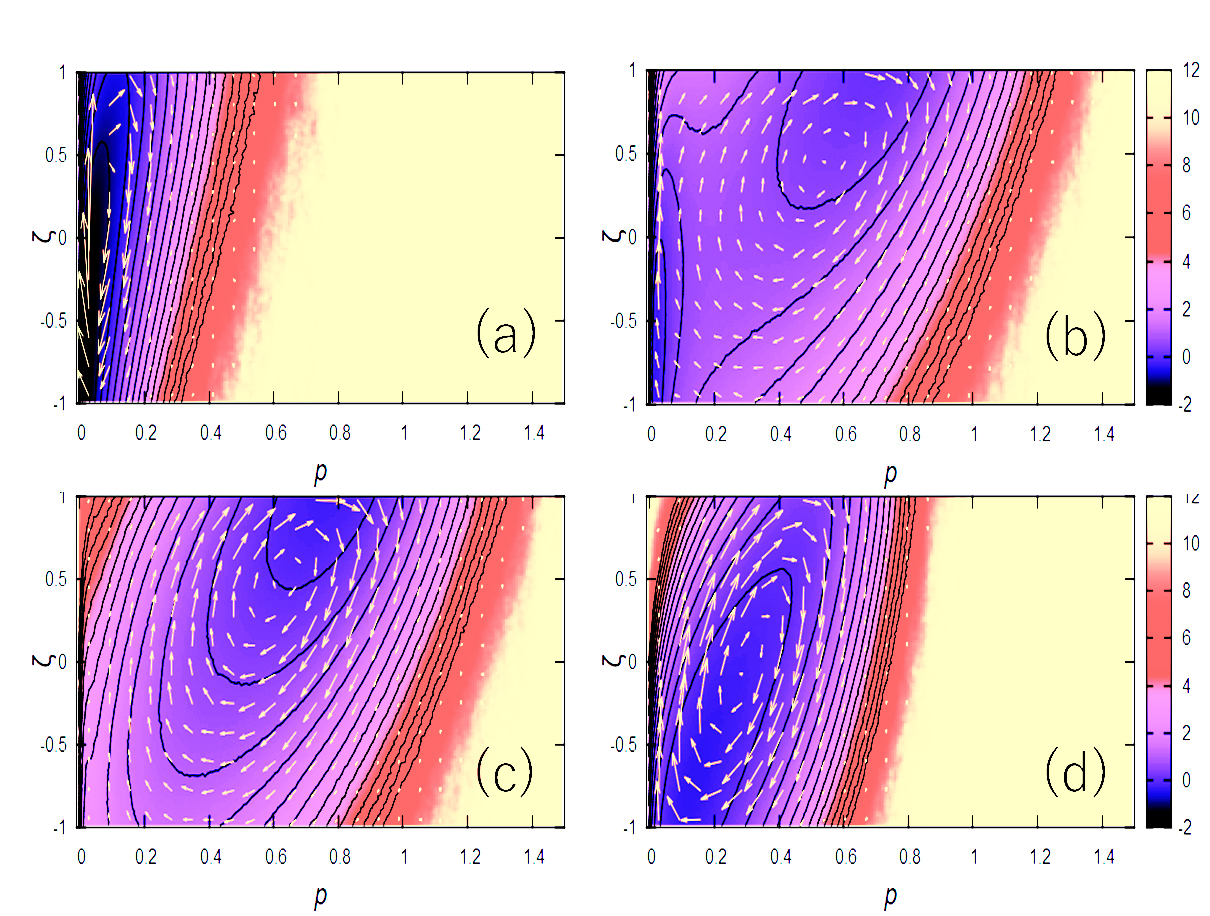}
\caption{Landscape $U(p,\zeta)$ and probability flux ${\bf J}(p,\zeta)$ calculated on the 2D plane of the protein concentration $p$ and the histone state $\zeta$ in the adiabatic approximation of TF-binding/unbinding. $U$ is shown with contours and ${\bf J}$ with yellow arrows. (a--c) Self-activating and (d) self-repressing genes. (a) $h_0/f=2$, (b) $h_0/f=20$, (c) $h_0/f=200$, and (d) $h_0/f=2$. The rate parameters are scaled in units of $k$ by setting $k=1$; 
$g_{11}=1$, $g_{-11}=g_{10}=0.2$, 
$g_{-10}=g_{1-1}=g_{-1-1}=0$, 
$\Delta r_{10}=\Delta r_{0-1}=1$, 
$\Delta r_{01}=\Delta r_{-10}=0$ for (a)--(c) and 
$g_{-11}=1$, $g_{11}=g_{-10}=0.2$, $g_{10}=g_{1-1}=g_{-1-1}=0$, $\Delta r_{10}=\Delta r_{0-1}=0$, and $\Delta r_{01}=\Delta r_{-10}=1$ for (d). For (a)--(d), $\Omega =100$, $c=20$, and $r_{10}^0=r_{0-1}^0=r_{01}^0=r_{-10}^0=1$. 
} 
\label{fig_landscape}
\end{center}
\end{figure}
%%%%%%%%%%%%%%%%%%%%%%%%%%%%%%%%%%%%%%%%%%%%%

Fig.~3 shows the calculated $U$ and ${\bf J}$ at the stationary state in the case of adiabatic TF-binding/unbinding. For an activator TF, when the TF-binding affinity is small (small $h_0/f$) (Fig.~3a), $U$ has a single basin at $p\approx 0$ and $\zeta\approx -1\sim 0$ (off state). When the binding affinity is intermediate (intermediate $h_0/f$) (Fig.~3b), $U$ has two coexisting basins in the off state and at $p\approx 0.7$ and $\zeta\approx 1$ (on state). When the binding affinity is large (large $h_0/f$) (Fig.~3c), a dominant basin is found at the on state. Thus, the binding affinity of the activator TF determines the distribution of stable states and works as a switch of gene states. When the TF is a repressor, we find a single basin at intermediate $p$ and $\zeta$ for a wide range of binding affinity (Fig.~3d).

In all the cases shown in Fig.~3, we find a flux ${\bf J}$ globally circulating around a basin or traversing between basins though the deterministic part of Eq.~9 is non-oscillatory; the oscillatory feature of the flow becomes evident through the stochastic  on-off switching fluctuations as a stochastic resonance effect. When the epigenetic effect is absent, the flux is diminished in the adiabatic limit \cite{Zhang2013,Chen2015}. However, here with nonadiabatic epigenetic histone dynamics, the circular flux is significant even in the limit of adiabatic TF-binding/unbinding because the timescales in $p$ and $\zeta$ are near to each other, so that the two processes are non-separable. The evident circulating flux suggests a temporal correlation between the histone modification and the gene activity change. By following the flux direction along the off-to-on (the on-to-off) path, the histone state first tends to become activating (repressing) followed by increase (decrease) in protein concentration.

%%%%%%%%%%%%%%%%%%%%%%%%%%%%%%%%%%%%%%%%%%%%%
\begin{figure}[t]
\begin{center}
\includegraphics[width=8.5cm]{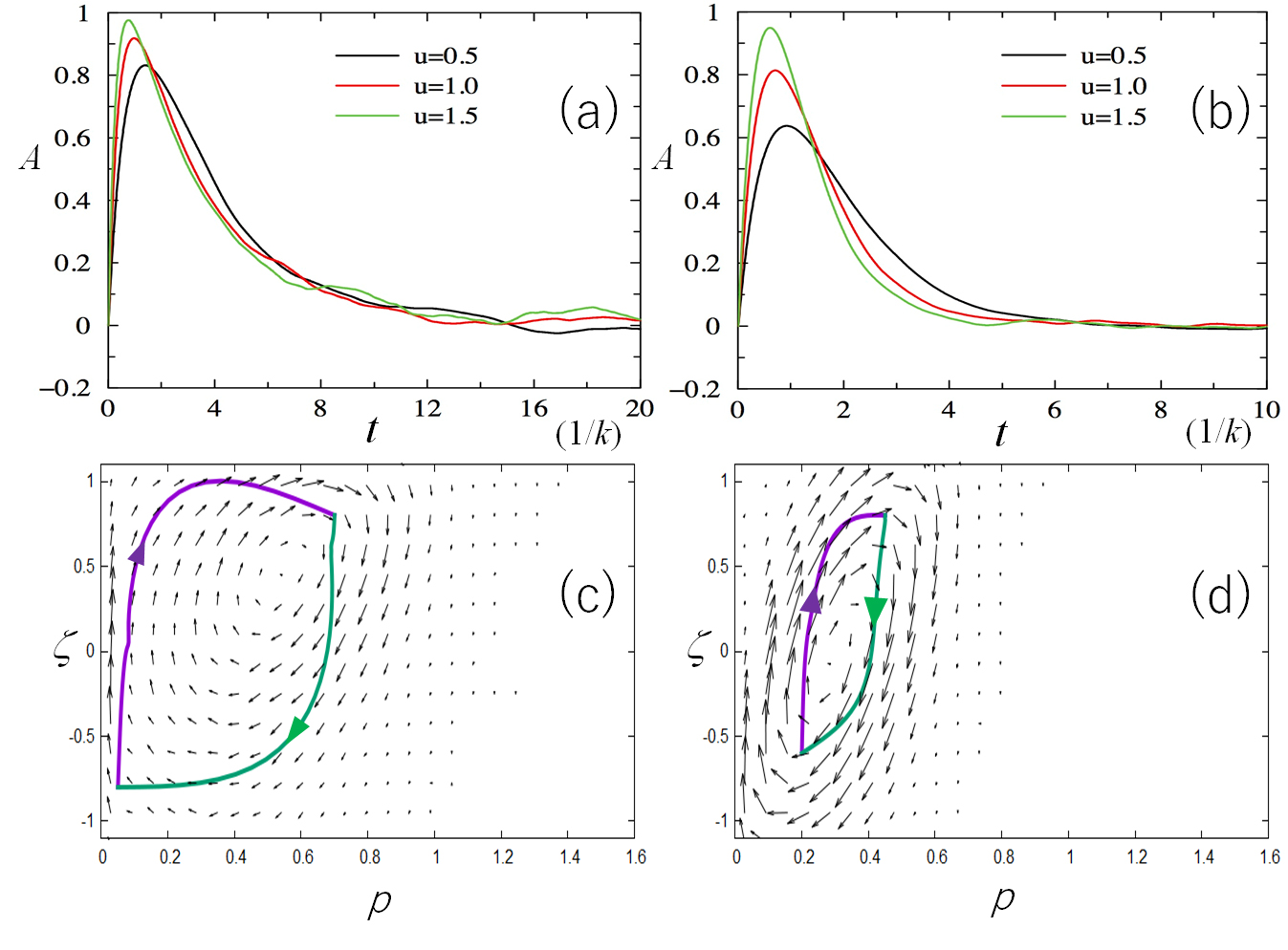}
\caption{Temporal correlation and optimal paths in self-regulating genes. (a,~c) Self-activating and (b, d) self-repressing genes. (a,~b) The normalized difference between two-time cross correlations, $A(t)$, is plotted as a function of $t$ in units of $1/k$. (c,~d) The optimal paths for the on-to-off (green) and off-to-on (purple) directions are superposed on the flux ${\bf J}$ on the 2D plane of the protein concentration $p$ and the histone state $\zeta$. Calculated using the adiabatic approximation of TF binding/unbinding. The rates of the histone state change are scaled by the parameter $u$ as $r_{10}^0=r_{0-1}^0=r_{01}^0=r_{-10}^0=u$ with $\Delta r_{10}=\Delta r_{0-1}=u$ in (a, c) or $\Delta r_{01}=\Delta r_{-10}=u$ in (b, d). In (a) and (b), $A(t)$ is plotted with $u=1.5$ (green), $u=1$ (red), and $u=0.5$ (black). In (c) and (d), $u=1$. The other parameters of (a, c) are same as in Fig.~3b and those of (b, d) are same as in Fig.~3d.
} 
\label{fig_pathway}
\end{center}
\end{figure}
%%%%%%%%%%%%%%%%%%%%%%%%%%%%%%%%%%%%%%%%%%%%%

This temporal correlation is confirmed by calculating the normalized difference between the two-time cross correlations, 
\begin{eqnarray}
A(t) =&&  \left[ \left<\delta\zeta(\tau)\delta p(\tau+t)\right> \right. \nonumber \\
&& \left. -\left<\delta p(\tau)\delta\zeta(\tau+t)\right>\right]/|\left<\delta \zeta(\tau) \delta p(\tau)\right>|,
\end{eqnarray}
with $\delta\zeta=\zeta -\left<\zeta\right>$ and $\delta p=p -\left<p\right>$, where $\left<\cdots\right>$ is the average over $\tau$ and the calculated trajectories. We can write $A(t) \propto \left< \det\left[{\bf q}(\tau), {\bf q}(\tau+t)\right]\right>$ with a 2D vector ${\bf q}(\tau)=(\zeta(\tau),p(\tau))$. A positive value of $A(t)$ at $t>0$ implies that the increase (decrease) in $\zeta$ tends to increase (decrease) $p$ at a later time $t$. For activator (Fig.~4a) and repressor (Fig.~4b) TFs, $A(t)$ is plotted for various values of $r_{ss'}^0/k=u$, showing that $A(t)$ has a positive-valued peak at $t_u =1/u\sim 0.5/u$. We find that the peak is evident even when the histone dynamics are as slow as $u<1$, which indicates that the prior change in the histone state to the gene activity is not owing to the faster rate of reactions in histones but is due to the circular flux generated by the nonadiabatic dynamics of histones. The delayed influence of $\zeta$ on $p$ should lead to the different on-to-off and off-to-on paths, inducing hysteresis in the switching dynamics.

The hysteresis is shown by calculating the optimal paths of transitions. An optimal path is obtained by first setting its start and end points and then minimizing the effective action in the path-integral formalism of kinetics connecting those points \cite{Aurell2002,Roma2005,Wang2010,Wang2011,Feng2014,Zhang2014b,Sood2020}. Figs.~4c and 4d show the paths calculated by the simulated annealing of the action with the algorithm of \cite{Wang2011}. Thus obtained off-to-on and on-to-off paths are indeed different from each other, both of which are consistent in their route orientations with the circulating probability flux. Because equilibrium kinetic paths should pass through the same saddle point of the landscape in both directions without showing the hysteresis, the calculated hysteresis is a manifestation of the nonequilibrium feature, and the area formed by the loop of paths gives a measure of the heat dissipation \cite{Feng2011b}.

%%%%%%%%%%%%%%%%%%%%%%%%%%%%%%%%%%%%%%%%%%%%%
\begin{figure}[tb]
\begin{center}
\includegraphics[width=8.5cm]{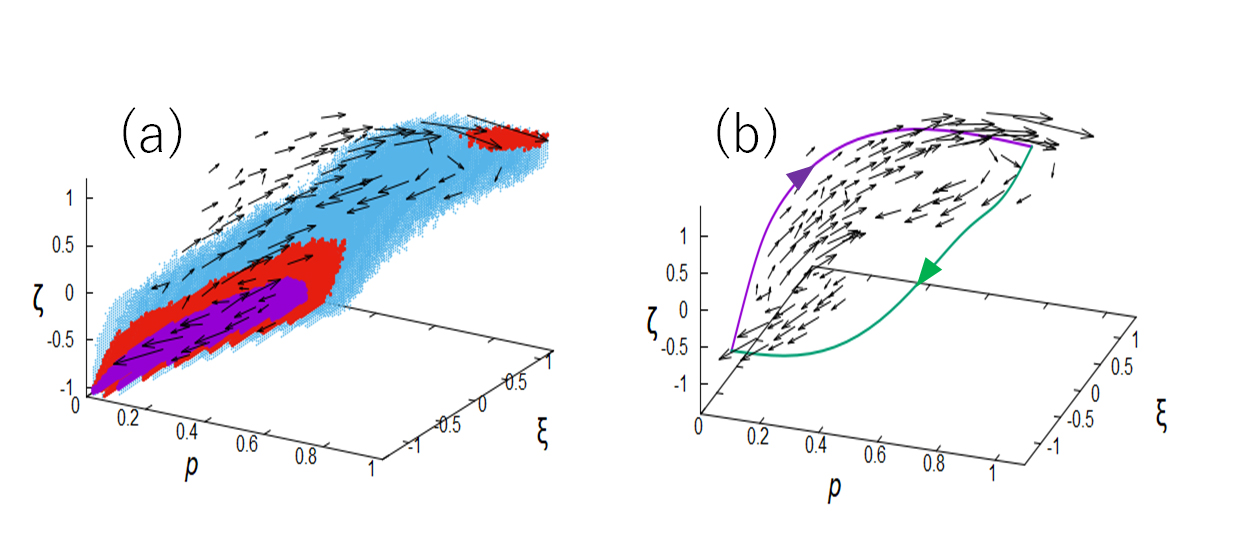}
\caption{Landscape, probability flux, and optimal paths of a self-activating gene calculated in the 3D space of the protein concentration $p$, the TF binding status $\xi$, and the histone state $\zeta$, in the nonadiabatic kinetics. (a) Flux ${\bf J}$ is superposed on the landscape $U$ with $1<U\le 2$ (light blue), $-0.5<U\le 1$ (red) and $U\le -0.5$ (purple). We find a gene-off state at $p\approx 0$, $\xi \approx -1$, and $\zeta \approx 0\sim -1$ and a gene-on state at $p\approx\xi \approx \zeta\approx 1$. (b) Flux and optimal paths of on-to-off (green) and off-to-on (purple) directions. The rate parameters are scaled in units of $k$ with $h_0=2$ and $f=0.1$. The other parameters are same as in Fig.~3b.
} 
\label{fig_landscape3d}
\end{center}
\end{figure}
%%%%%%%%%%%%%%%%%%%%%%%%%%%%%%%%%%%%%%%%%%%%%

When the TF binding/unbinding is slow, we need to solve Eq.~9 nonadiabatically. Slow nonadiabatic binding/unbinding were examined recently by tuning the binding rate in bacterial cells experimentally \cite{Jiang2019,Fang2018}. Fig.~5a shows the landscape and flux in such a nonadiabatic binding/unbinding case with the intermediate binding affinity of the activator TF. We find two basins for the on and off states and a distinct circular flux between them. The qualitative feature is same as in the adiabatic TF-binding/unbinding case; however, here, we find the correlated binding/unbinding behavior with the histone-state change, which should generate hysteresis in the 3D space. This hysteresis can be found in the calculated optimal paths in the 3D space (Fig.~5b).

\section{Discussion}

The present flux-landscape analyses provided a new view that the nonadiabatic circular flux generates nontrivial temporal correlation, hysteresis, and dissipation in eukaryotic gene switching. 
These analyses provide a clue to resolving the `chicken-and-egg' argument on the causality between the histone state and TF binding. The landscape analyses showed that the stability of each histone state is determined by the binding affinity of TF, which is in accord with the general belief that the specific TF binding is the cause and the histone-state change is the result \cite{Ptashne2013}. However, unexpectedly, the histone state tends to change in self-regulating gene circuits prior to the change in the amount of TF, which induces hysteresis in the switching dynamics; the histone-state fluctuation can be a trigger for switching the feedback loop. This temporal correlation should be tested experimentally by analyzing $A(t)$ in single-cell observations. This type of experiments should be possible when the amount of the product TF is measured by a co-expressing fluorescent protein and the live-cell histone state is monitored simultaneously  by  the technique of  fluorescently labeled specific antigen binding  \cite{Hayashi-Takanaka2011,Sato2019}. The flux structure and timescales should be also  tested experimentally  by examining the irreversibility in temporal correlations \cite{Liu2020}.

A possible test of the present model is to quantitatively monitor the response of somatic cells to the incorporation of exogenous genes  such as Yamanaka factors, which include  {\it Sox2} and {\it Oct4}   \cite{Takahashi2006}. By introducing Yamanaka factors, the differentiated somatic cells can turn into the pluripotent cells. This reprogramming  of cells may start from the binding of exogenous pioneer factors, Sox2 and Oct4, to the loci of endogenous genes, {\it Sox2} and {\it Oct4}. By writing the concentration of exogenous Sox2 and Oct4 proteins as $p_{ex}$, the total concentration of a Sox2-Oct4 heterodimer TF should be approximately proportional to $(p+p_{ex})^2$ in the present model; hence, the TF binding rate should become $h=h_0(p+p_{ex})^2$. Shown in Fig.~6 are the landscapes and fluxes for the small (Fig.~6a) and large (Fig.~6b) values of  $p_{ex}$ with the parameters of Fig.~3a, with which the landscape is dominated by the off-state when $p_{ex}=0$.  We find that the landscape is shifted upon introduction of exogenous factors with $p_{ex}>0$ to have a basin in the on-state, but the flux structure remains similar to that in Fig.~3, implying the strong tendency of the histone-state change before the changes in the activity of the endogenous genes when the endogenous gene is modified from the off-state.

%%%%%%%%%%%%%%%%%%%%%%%%%%%%%%%%%%%%%%%%%%%%%
\begin{figure}[t]
\begin{center}
\includegraphics[width=8.5cm]{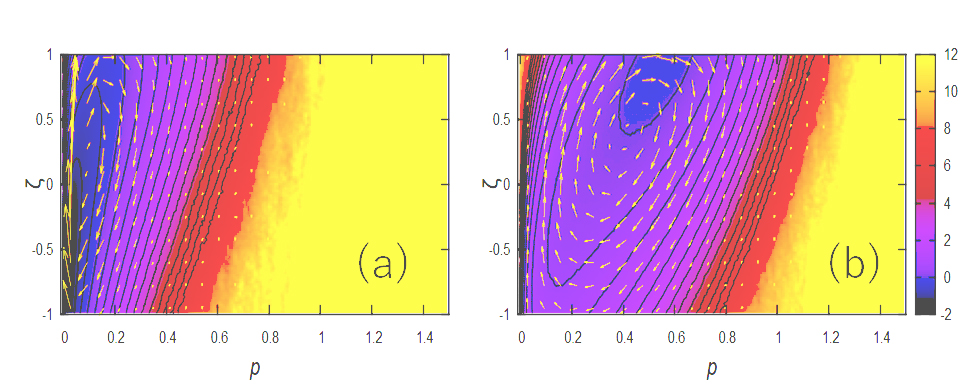}
\caption{Landscape $U(p,\zeta)$ and probability flux ${\bf J}(p,\zeta)$ with the exogenous TF contribution $p_{ex}$.
(a) $p_{ex}=0.1$ and (b) $p_{ex}=0.6$. Calculated on the 2D plane of the protein concentration $p$ and the histone state $\zeta$ in the adiabatic approximation of TF-binding/unbinding. The parameters are the same as in Fig.~3a. 
} 
\label{fig_landscape}
\end{center}
\end{figure}
%%%%%%%%%%%%%%%%%%%%%%%%%%%%%%%%%%%%%%%%%%%%%

Finally,  the flux-landscape method should be applicable to  problems of other epigenetic degrees of freedom. For example, a Monte-Carlo simulation of a gene network suggested that formation/dissolution of a super-enhancer of {\it Nanog} induces a large fluctuation in  mESCs \cite{Sasai2013}. It is intriguing to develop a method to explain the degree of freedom of super-enhancer formation/dissolution and the associated large-scale chromatin structural change by extending the present scheme. Thus, the flux-landscape approach should provide physical insights into various problems of nonadiabatic stochastic switching.

\section*{Acknowledgements}
B. B. and M. S. thank Dr. S. S. Ashwin for fruitful discussions. This work was supported by JST-CREST Grant JPMJCR15G2, the Riken Pioneering Project, and JSPS-KAKENHI Grants JP19H01860, 19H05258 and 20H05530. J. W. thanks NSF-PHY-76066 for supports.

%%%%%%%%%%%%%%%%%%%%%%%%%%%%%%%%%%%%%%%%%%%
% Bibliography
\bibliography{Bhaswati}

\end{document}